\documentclass[global,twocolumn,hyperref]{svjour-arXiv}

\usepackage{graphics}
\usepackage{here}
\usepackage{textgreek}
\usepackage{cite}

\begin{document}
\title{Optical Second-harmonic Images of Sacran Megamolecule Aggregates}

\author{\and{Yue Zhao}\inst{1}\hbox{\href{http://orcid.org/0000-0002-8550-2020}{\includegraphics{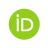}}} 
\and Khuat Thi Thu Hien\inst{1} 
\and Goro Mizutani\inst{1}\thanks{\emph{Goro Mizutani:} mizutani@jaist.ac.jp}\hbox{\href{http://orcid.org/0000-0002-4534-9359}{\includegraphics{orcid.eps}}}
\and Harvey N. Rutt\inst{2}
\and Kittima Amornwachirabodee\inst{1} 
\and Maiko Okajima\inst{1} \and Tatsuo Kaneko\inst{1} 
}                     
%
%

\institute{School of Materials Science, Japan Advanced Institute of Science and Technology, Asahidai 1-1 Nomi, 923-1292, Japan \and School of Electronic and Computer Science, University of Southampton, SO17 1BJ, UK}

\date{
{\it J. Opt. Soc. Am. A}, {\bf 34,} 146-152 (2017).  DOI: 10.1364/JOSAA.34.000146}
\maketitle
\begin{abstract}
\bf{
We have detected a second-order nonlinear optical response from aggregates of the ampholytic megamolecular polysaccharide sacran extracted from cyanobacterial biomaterials, by using optical second harmonic generation (SHG) microscopy. The SHG images of sacran cotton-like lump, fibers, and cast films showed SHG intensity microspots of several tens of micrometers in size. The dependence of the SHG spot intensity on an excitation light polarization angle was observed to illustrate sacran molecular orientation in these microdomains. We also observed SHG signals around a special region of the cast film edges of sacran. These results show that sacran megamolecules aggregate in several different ways.
}
\end{abstract}

\section{Introduction}
Cyanobacterial cells can be regarded as efficient microreactors for conversion of carbondioxide, water, and nitrogen, into other useful, renewable molecules under light energy. The cyanobacteria produce exopolysaccharides (EPS) attracting attention of researchers in various fields from microbiology to physical chemistry not only because it is water soluble and easily recovered from liquid cultures\cite{k1,k2,k3,k4} but also because some of EPS have functions such as antivirus\cite{k5,k6}, heavy metal adsorption\cite{k7,3,4,5,7,8}, water retention ability\cite{k8,2}.  According to a review, more than 100 cyanobacteria have been reported to synthesize a large amount of EPS\cite{k4}.   The molecular mass of EPSs ranges from 200 to 1 600 kDa, and more than 75 \% of those so far characterized are heteropolysaccharides consisting of six or more different kinds of monosaccharides. In recent years, we developed the EPS, sacran, of {\sl Aphanothece sacrum }which has been cultured in a traditional style for more than 100 years in Japan.

\begin{figure}[h]
\begin{minipage}[t]{8.5cm}
\resizebox{1\textwidth}{!}{  \includegraphics{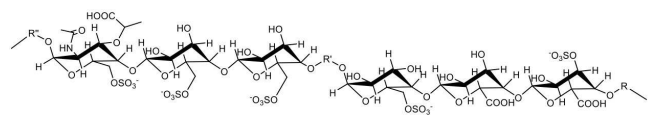}}
\end{minipage}%
\caption{A partial structure formula of sacran.\cite{3}}
\label{structure_formula}  
\end{figure}

Sacran is a non-crystalline megamolecular polysaccharide with absolute molecular weight ranging over 10 million g/mol\cite{1,2}.  Its molecular weight is the largest among the ones ever reported so far. A partial structure of sacran molecule is shown in the Fig. \ref{structure_formula}. The non-crystallinity is derived from a very complex structure consisting of at least ten kinds of constituent sugar residues such as uronic acid, sulfated and aminated sugars.   Sacran can thus be regarded as an amphoteric electrolyte. Owing to the megamolecular polyelectrolyte structure having ca. 100 000 anions and 1 000 cations in a chain, sacran shows some unique functions, 1) ultra-high water-retention capacities of 6100 times its dry weight for pure water and of 2600 times for saline.\cite{1} 2) Sacran forms a rod-like organization with a length over 10 \textmu m in the presence of sodium chloride. 3) Sacran can efficiently adsorb trivalent metal ions such as neodymium\cite{3,4} and indium ions\cite{5,6,7,8,9} and other divalent metal ions of large atomic number.\cite{10}

When the concentration of sacran in aqueous solution is lower than the chain overlap concentration (0.004 wt\%), sacran chains adopt a spherical conformation. On the other hand, the conformation transforms into a rigid-rod one at increased concentrations of over 0.1 wt\%, and at even larger concentration, it exhibits a liquid crystalline phase.\cite{5,10} However, the structure and various physical properties of oriented sacran aggregates have not yet been elucidated. More generally, the cyanobacterial polysaccharides have not been well investigated, although the cyanobacterium is important as the possible origin of plant chloroplasts and is regarded as an eco-friendly microreactor to produce various organic materials by not only carbon- but also nitrogen-fixation under an extremely oligotrophic environment. Although some structures have already been observed with AFM\cite{10}, SEM\cite{SEM1,SEM2} and TEM\cite{4} as imaging information, to date, there is lack of understanding of their complicated asymmetric structure. 

Hence the main purpose of this study is to observe and analyze the asymmetric structure and orientation of megamolecule sacran aggregates by using an SHG (Second Harmonic Generation) microscope. Sacran contains a large amount of chiral structures based on sugar chains and it should allow SHG. It is also expected that SHG will be enhanced by the anisotropically oriented chain structure of sacran.

Optical second harmonic generation (SHG) is one of the second-order nonlinear optical phenomena. When light pulses pass through non-centrosymmetric materials such as chiral molecules or non-centrosymmetric structural units, they generate frequency doubled pulses.  It is well known that non-centrosymmetric crystals have non-zero second-order nonlinear susceptibilities\cite{11} and some macroscopic biomaterials with ordered structure have an extremely high second-order nonlinear susceptibility\cite{12,13,14,15,16,17,18,19,20,21,22,23,24,25,26,27,28,29,30,31,32,33,34}.  Based on this property, SHG microscopes can selectively visualize ordered structures on the basis of such a difference in SHG activity.\cite{12} Chirality enhances the second-order nonlinear optical effect in cooperation with macroscopic order\cite{t2}. For example, collagen in the living body has a triple helix structure of polypeptide chains aligned in the same direction, called procollagen, and this asymmetric structure allows SHG \cite{13}. Thus the SHG microscopic image\cite{12,14,15,16} is useful in histochemical assay of pathological and medical applications \cite{17,18,19,20,21,22,23,24,25}. SHG microscopic images were also observed from fluorescent protein (chromophores)\cite{26} and some structural proteins\cite{27} in the living body such as tubulin\cite{28} and myosin.\cite{15,29} Cellulose from outside of microbial cells also generate SHG. Longer cellulose fibers generate stronger SHG signals.\cite{30,31} Since starch has a higher-order asymmetric structure on the macroscopic scale, its SHG microscopic image was observed very strongly.\cite{32,33,34} Selective observation of asymmetric molecules and nano-structure can also be performed for liquid crystals.\cite{35,36} The SHG microscope has an advantage over the fluorescence microscope that it is not necessary to dye the target sample, and thus photobleaching and cytotoxic by-products do not arise.\cite{37} This property is useful in order to observe living cells and nerves.\cite{38} In this study, we use a home-made femtosecond laser SHG microscope to observe images of sacran aggregates. 

\section{Methods and Materials}
\subsection{Experimental Setup}

\begin{figure}[h]
\begin{minipage}[t]{8.5cm}
\resizebox{1\textwidth}{!}{  \includegraphics{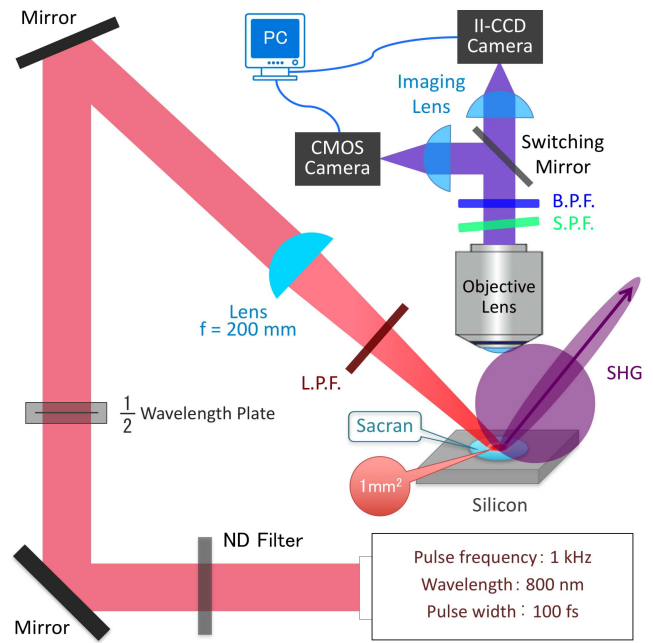}}
\end{minipage}%
\caption{SHG microscopy system. A femtosecond laser was used as the light source. The laser beam was focused loosely on the sample from an off axis direction of the collection optics. Scattered SHG light from the sample was imaged by an objective lens and detected by an image intensifier - charge coupled device (II-CCD) camera (HAMAMATSU, PMA- 100). We used a CMOS camera (Lumenera corporation, Lu135M) for aiming at a target. The sacran samples were put on silicon wafer substrates. When 2PEF images were observed, we only replaced the band pass filter of 400 nm into that of 438 nm.}
\label{setup_ver_6_5}  
\end{figure}

The light source of the experiment was a regenerative amplifier excited by a mode-locked Ti:sapphire laser, with a repetition frequency of 1 kHz, a pulse width of about 100 fs, and wavelength of 800 nm. The laser power at the sample position was controlled by an ND (Neutral Density) filter. The laser beam was focused on the sample by a lens with a focal length of 200 mm. Between the focusing lens and the sample, we put a long wavelength pass filter (L.P.F.), to remove light of wavelength shorter than 780 nm, and pass the 800 nm beam. The SHG from the sample was observed in a scattering geometry. In order to make the irradiated area on the sample about 6 mm$^2$, the sample was placed at 5 mm off the focal point of the lens ($f =$ 200 mm). The intensity distribution of the incident beam cross-section was roughly Gaussian. However, the incident angle with respect to the normal to the sample stage was 60$^\circ$, and hence the irradiated spot shape on the sample was elliptical. As imaging optics a commercial microscope OLYMPUS BX60 was used. SHG light from the sample passed through the objective lens, became a parallel ray, passed through a short wavelength pass filter (S.P.F.) passing 350 - 785 nm and rejecting 800 nm wavelength, and finally was selected by a ‘Semrock’ band pass filter (B.P.F.) FF01-395/11 with the transmittance over 90\% at 388.4 - 402.2 nm and 92.9\% at 400 nm. Sacran also showed 2-photon excitation fluorescence (2PEF) in the same wavelength region as SHG. In order to check the image of the 2PEF, we used an alternative ‘Semrock’ band pass filter FF02-438/24 with a center wavelength of 438 nm. The short wavelength pass filter was placed at a tilt angle of 5$^\circ$ with respect to parallel rays to eliminate ghost signals due to interference. The image was observed using a photon counting function of an image intensified - charge coupled device (II-CCD) camera (HAMAMATSU, PMA-100).  We used a CMOS camera (Lumenera corporation, Lu135M) for aiming at a target. The spatial resolution of the microscope is decided by the chip size of the II-CCD camera, i.e. 11 \textmu m $\times$  13 \textmu m, and was 2.6 \textmu m for magnification  $\times$ 5 (NA=0.15) and  0.65 \textmu m for magnification  $\times$ 20 (NA=0.46).  In this experiment, the maximum excitation light power on the sample is 8.6 mW.  When calculated with an irradiation area of 1 mm$^2$, the power density per unit area is 8.6 mW/mm$^2$.  Since the repetition frequency is 1 kHz, the excitation light energy density of one pulse is 8.6 \textmu J/mm$^2$.

The energy of the amplified pulse (1 kHz) is 8$\times$10$^4$ times the seed light (80 MHz). If this pulse enters the objective lens, it will cause various nonlinear optical effects, cause broadband white light, and make SHG light indistinguishable. The forward or backward collection configuration could not be adopted because of this effect. 



\subsection{Preparation of the Sample}
The extraction of sacran from {\sl Aphanothece sacrum} was done as  follows:

(1) Frozen {\sl Aphanothece sacrum} was re-melted to break the cell membrane and to remove water-soluble substance such as phycobiliprotein eluted from cell by washing with water.

(2) Fat-soluble matters such as lipids, chlorophyll, and carotenoid-based dye were washed away by isopropa-nol.

(3) The washed {\sl Aphanothece sacrum} samples were put into 0.1N-NaOH aqueous solution and kept at 80 $^\circ$C with vigorous stirring for 5 hours, to decompose biomacromolecules such as proteins and DNA. Polysaccharides degrade more slowly under alkaine conditions. 

(4) The solid impurities were removed by filtration, and the filtrate was neutralized to approximately pH 7-8 using HCl aqueous solution.

(5) The neutralized solution was poured into an aqueous solvent containing isopropanol and water with the ratio of 7:3, to precipitate the high-molecular-weight poly-saccharide whilst keeping low-molecular-weight matters soluble.

(6) The polysaccharide obtained by filtration was dissolved in water and recovered again using the water/isopropanol mixture for futher purification. The purification was made three times in total to get white fibrils. The polysaccharide fibrils were dried in an oven at 60 $^\circ$C for 2 hours, and the sacran ‘cotton’ was obtained (Fig. \ref{lump_and_fiber}(a)).

(7) The pure sacran ‘cotton’ was dissolved in distilled water, and the solution was dropped directly on a silicon substrate, to dry spontaneously at room temperature to form a film. The films thus obtained are shown in Figs. \ref{filter}(a) and (e) and in Fig. \ref{film}(a).

In this experiment, silicon wafers were used as the substrates for SHG measurement of sacran. This is because silicon wafers generate negligible SHG. 

\section{Results}
\subsection{SHG Microscopic Image of Pure Sacran}

\begin{figure}[h]
\begin{minipage}[t]{8.5cm}
\resizebox{1\textwidth}{!}{  \includegraphics{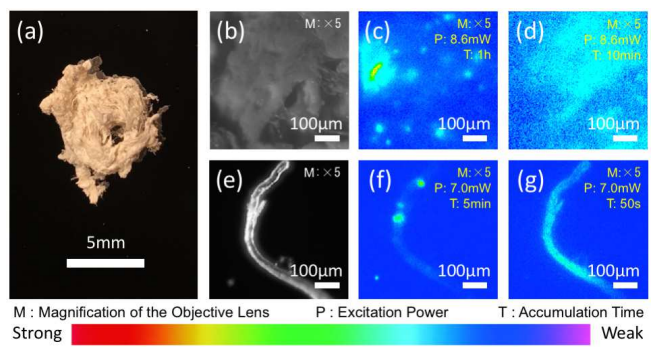}}
\end{minipage}%
 \caption{(a) Cotton-like high purity sacran extracted and fiberized from {\it Aphanothece sacrum}. (b) Microscopic image of cotton-like pure sacran using white light illumination and a CMOS camera. (c) SHG image observed by using a band pass filter of 400 nm wavelength, and (d) 2PEF image observed by using a band pass filter of 438 nm wavelength. (e) Micro- scopic image of a sacran fiber taken from the cotton-like sacran lump. (f) Corresponding SHG image, and (g) 2PEF image.} 
\label{lump_and_fiber}
\end{figure}

\begin{figure}[h]
\begin{minipage}[t]{8.5cm}
\resizebox{1\textwidth}{!}{  \includegraphics{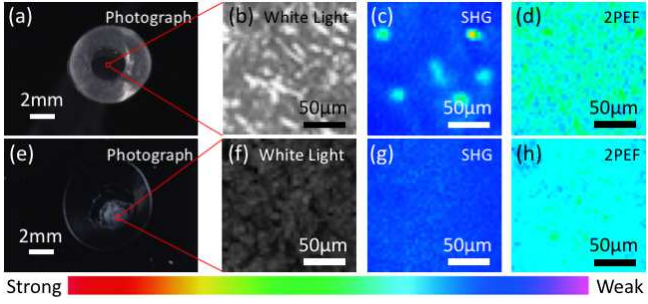}}
\end{minipage}%
 \caption{Images of unfiltered (a-d) and filtered (e-h) cast films made from a sacran aqueous solution with concentration of 0.5\%. The film thicknesses are approximately 5 \textmu m thick. (a) Photograph of a cast film dried at room temperature from a drop of the sacran aqueous solution on a silicon substrate. (b) Expanded image of the area in a red frame in (a) using white light illumination and a CMOS camera, (c) and (d) Corresponding SHG microscopic intensity image and the 2PEF image, respectively. (e) Photograph of a cast film dried at room temperature from a drop of the sacran aqueous solution filtered by a membrane filter of 0.45 \textmu m pore size, on a silicon substrate. (f) Expanded image of the area in a red frame in (e) using white light illumination and a CMOS camera, (g) and (h) Corresponding SHG and 2PEF images.  The magnification of the objective lens was $\times$5 (NA = 0.15). For (c), (d), (g), and (h) the incident light wavelength was 800 nm, and the excitation power was 6.6 mW. The integration time of (c) and (g) was 5 min, and that of (d) and (h) was 50 s.}
\label{filter}
\end{figure}

Figure \ref{lump_and_fiber}(b) is a linear optical microscopic image of a pure sacran ‘cotton’ lump with white light illumination.  
Figure \ref{lump_and_fiber}(c) is an image illuminated by the 800 nm light pulses observed by the II-CCD camera with band pass filter of 400 nm wavelength set before the camera.  Figure \ref{lump_and_fiber}(d) is an image observed after the band pass filter was replaced to 438 nm wavelength.  There are two possible candidate origins for the 400 nm signal in Fig. \ref{lump_and_fiber}(c). One is SHG signal and another is luminescence. In the latter case the sample was excited by a multi-photon transition, and emitted luminescent pulses at the photon energy lower than the multi-photon energy. If the signal in Fig. \ref{lump_and_fiber}(c) is dominated by the latter, just like in Fig. \ref{lump_and_fiber}(d), the observation wavelengths of 400 nm and 438 nm are not expected to show a big difference, so that the images in Figs. \ref{lump_and_fiber}(b) and (c) should be similar to each other. However, the spatial distributions of the signal are clearly very different between in Figs. \ref{lump_and_fiber}(b) and (c). Thus, it is judged that the signal in Fig. \ref{lump_and_fiber}(c) is SHG, while the signal in Fig. \ref{lump_and_fiber}(d) is 2PEF. This result indicates that SHG signal and an SHG microscopic image of the ultra-macromolecules sacran aggregates have been observed for the first time. In Fig. \ref{lump_and_fiber}(c), the SHG intensity is distributed in spots with a size of about tens of micrometers. 

When we tilted the cotton-like sacran by several different angles while observing the SHG image, the relative intensity between the SHG spots changed.  This shows that the SHG light comes out in a different angle from each spot.  In order to clarify the origin of the SHG spots in Fig. \ref{lump_and_fiber}(c), a small sacran fiber was picked out from pure sacran ‘cotton’ lump and was observed as shown in Fig. \ref{lump_and_fiber}(f). In Fig. \ref{lump_and_fiber}(f), one sees granular SHG spots similar to the ones in Fig. \ref{lump_and_fiber}(c) with a size of tens of micrometers. The SHG spots are locally distributed on the sacran fibers. 

We have also observed two kinds of films with a thickness of approximately 5 \textmu m with and without filtering the solution by a membrane filter before dropping it on the Si substrates. In the SHG image (Fig. \ref{filter}(c)) of the film dried from a sacran aqueous solution directly dropped on a silicon substrate, strong intensity spots were observed similar to those of the pure sacran ‘cotton’ lump (Fig. \ref{lump_and_fiber}(c)) and fiber (Fig. \ref{lump_and_fiber}(f)). On the other hand, a film dried from the sacran aqueous solution dropped on a silicon substrate via a membrane filter with the hole diameter of 0.45 \textmu m, did not show granular SHG intensity spots as seen in Fig. \ref{filter}(g).

\subsection{Incident Polarization Dependence of the SHG Spots}

\begin{figure}[h]
\begin{minipage}[t]{8.5cm}
\resizebox{1\textwidth}{!}{  \includegraphics{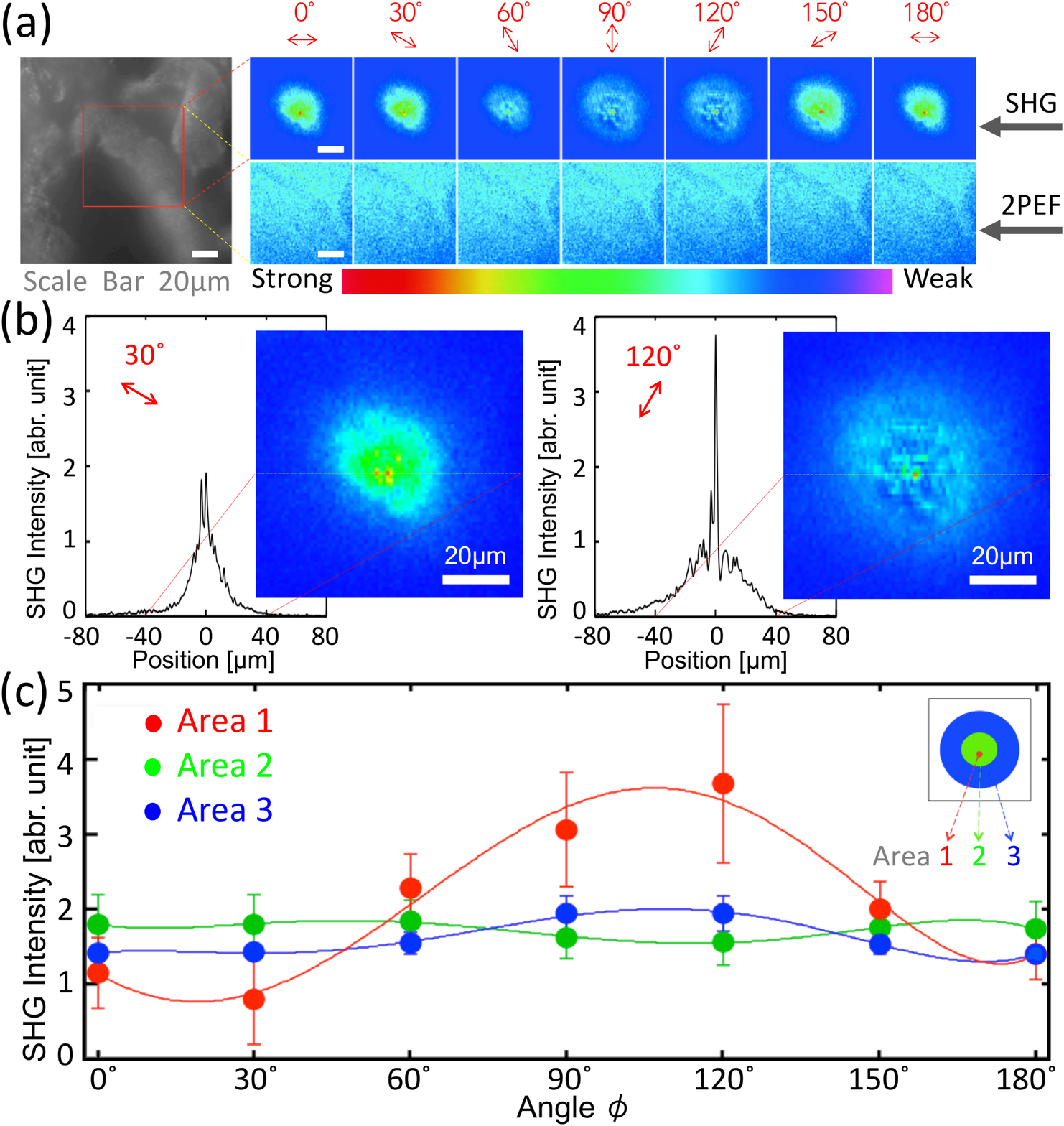}}
\end{minipage}%
\caption{Incident polarization dependence of the SHG intensity and the 2PEF images of pure sacran at one SHG spot. (a) The left panel is a linear microscopic image obtained with a CMOS camera and the right upper row images are the SHG microscopic images, and the right lower row ones are 2PEF microscopic images. Incident light wavelength was 800 nm, the average excitation power was 6.8 mW, the integration time for the SHG and 2PEF image observations were 300 s and 50 s, respectively. In the image observation the polarization was not specified. The magnification of the objective lens was $\times$ 20 with NA = 0.46. (b) Intensity profiles in the strongest SHG spot for 30$^\circ$ and 120$^\circ$ input polarization angle. The intensity was collected digitally from the image file at the positions on the yellow dashed line shown in the inset images. The collection width of the profile line was one pixel. (c) The SHG intensity variations of incident polarization in Areas 1, 2 and 3.
}
\label{polarization}
\end{figure}

Figure \ref{polarization}(a) shows the excitation light polarization dependence of an SHG intensity spot from the pure sacran ‘cotton’ lump. Incident polarization angle is shown at the top and the angle is defined as 0$^\circ$ when the incident light electric field is directed parallel to the sample stage. The intensity of the SHG spots clearly depends on the input polarization angle. On the other hand, 2PEF has nearly constant intensity for all the incident polarization angles. The excitation light polarization dependence of the SHG intensity images follows the selection rule of SHG. It shows that sacran molecules were well oriented in one direction in the spot area. 

A careful look at the SHG spot in Fig. \ref{polarization}(a) shows that the SHG intensity is stronger in the 2 \textmu m diameter area at the center. In addition, the total size of the SHG spot changes as a function of the excitation polarization angle in Figs. \ref{polarization}(a). In Fig. \ref{polarization}(a) at 0$^\circ$ and 30$^\circ$, the diameter of the SHG spots is about 30 \textmu m, while at 90$^\circ$ and 120$^\circ$, the diameter of the SHG spots is about 60 \textmu m. Figure \ref{polarization}(c) shows the SHG intensity variations of incident polarization in the three areas defined in the inset of Fig. \ref{polarization}(c). The SHG intensity of Areas 1 and 3 is the strongest at the same polarization angle of 120$^\circ$. The strongest polarization angles of Area 2 is different from that of Areas 1 and 3 by 90$^\circ$. We suggest that this strong sacran SHG region is made up from three regions with diameters of 2 \textmu m (Area 1), 30 \textmu m (Area 2), and 60 \textmu m (Area 3). We could see a similar tendency in the image structures in other SHG spots.

\subsection{SHG Images of Sacran Thin Films}

\begin{figure}[h]
\begin{minipage}[t]{8.5cm}
\resizebox{1\textwidth}{!}{  \includegraphics{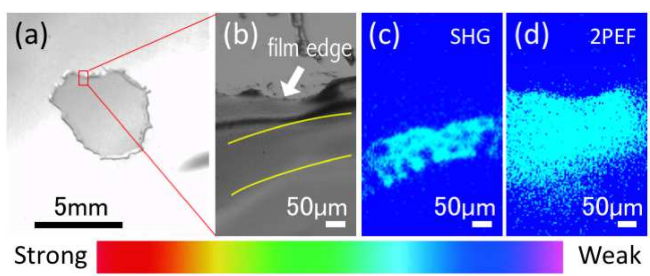}}
\end{minipage}%
 \caption{Linear and nonlinear images of a cast film made from sacran aqueous solution dropped on a silicon substrate after passing through a membrane filter of 0.45 \textmu m pore size. (a) Its photograph image, (b) Its microscopic image using a CMOS camera, (c) Its SHG image, and (d) Its 2PEF image. In the SHG image of (c) and 2PEF image of (d), the incident light wavelength was 800 nm, average excitation power was 7.0 mW, and accumulation time was 5 min and 50 s. The red square frame in (a) is the observed area of the microscopic images of (b). Yellow lines in microscopic images (b) indicate the area of strong SHG intensity in (c). The magnification of the objective lens was $\times$5 with NA = 0.15.}
\label{film}
\end{figure}

As seen in Fig. \ref{filter}(g), the SHG intensity at the center of the cast film prepared using the 0.45 \textmu m membrane filter was at the noise level. On the other hand, some areas near the edges of the same film showed SHG as shown in Fig. \ref{film}(c). This SHG was not like the spotty signals observed in Figs. \ref{lump_and_fiber}, \ref{filter} and \ref{polarization}, but had wider spread. It is observed for all the sacran films but only near their edge regions. On the other hand, the intensity distributions of 2PEF in Fig. \ref{film}(d) is almost similar to that of the excitation light. 2PEF is generated everywhere from sacran, and does not depend on the position in sacran. 

In Fig. \ref{film}(c), the SHG emitting area is not completely continuous, and some areas with no SHG of several tens of micrometer sizes are seen. In the linear image of Fig. \ref{film}(b) the yellow curves indicate the area of intense SHG emission, but one cannot see any particular visual difference of this area from the outer areas. The upper boundary of this SHG emitting area lies roughly parallel to the edge of the film with a distance of several tens of micrometers. The SHG intensity of the films was qualitatively weaker than that of sacran ‘cotton’ lump and sacran fibers.

In this study, the maximum excitation light power density was 8.6 mW/mm$^2$. In order to check the damage threshold, the excitation light power was raised up to 60 mW/mm$^2$ with the same optical setup, but damage such as burn marks of the sample could not be found by the CMOS camera.  Therefore, damage in the sample is considered to be negligible within the scope of this study.

\section{Discussion}

First, we discuss the origin of the SHG spots with the size of ten to several tens of micrometers observed for the pure sacran ‘cotton’ lump (Fig. \ref{lump_and_fiber}(c), Fig. \ref{polarization}(a)), the sacran fibers (Fig. \ref{lump_and_fiber}(f)), and the sacran films (Fig. \ref{filter}(c)).   In Fig. \ref{lump_and_fiber}(e) sacran is a fiber-like aggregate and the material density does not appear to change drastically around the parts showing SHG spots.  This indicates that sacran is distributed rather continuously along the fiber axis, while the SHG emission is very localized. In Fig. \ref{polarization} the SHG spot intensity from pure sacran depends strongly on the polarization of the incident light. The diameters of the SHG spots are approximately 30 \textmu m (Area 2) for 0$^\circ$ and 30$^\circ$, and 60 \textmu m (Area 3) for 90$^\circ$ and 120$^\circ$. At the center of the spots, the SHG area with 2 \textmu m (Area 1) diameter has different polarization dependence, and it is strongest at 120$^\circ$, and weakest for 30$^\circ$. Namely, these domains with different polarization dependence and different diameters of 2 \textmu m (Area 1), 30 \textmu m (Area 2) and 60 \textmu m (Area 3) are overlapped concentrically. 

When there is an electric field associated with an electric charge in the sacran aqueous solution, then it generates a nucleus of sacran molecule aggregates and the solution will change into the liquid crystal phase. In the case of this study, sacran cations,\cite{1} large sacran molecules or some impurities such as cell fragments may have triggered heterogeneous nucleation of the sacran molecules. As the solution dried, it became oversaturated and a liquid crystal domain was formed around the nucleus. The spotty SHG microscopic images in pure sacran ‘cotton’ lump or a sacran fiber show core and spherical shell structures (see Fig. \ref{polarization}(c)) and these core and shells have liquid crystal domains with their respective molecular orientation. Namely, SHG spots of 30 \textmu m size at 30$^\circ$ show the molecules in the same orientation direction in one liquid crystal domain. Those of 60 \textmu m size at 120$^\circ$ show molecules in another orientation of a liquid crystal domain. 

\begin{figure}[h]
\begin{minipage}[t]{8.5cm}
\resizebox{1\textwidth}{!}{  \includegraphics{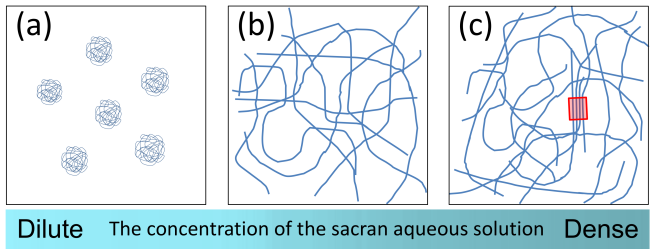}}
\end{minipage}%
\caption{(a) Schematic sacran molecules in a sacran aqueous solution at low concentration, (b) at higher concentration, and (c) at still higher concentration. As the sacran concentration increases, sacran molecules extend from spherical (a) to soft filaments (b).\cite{39} At more higher concentration there occurs a liquid crystal precursor due to impurities or large sacran (red frame) (c).}
\label{mechanism1}
\end{figure}

As one candidate origin of this molecular configuration, we suggest a model of Fig. \ref{mechanism1}. In the low-concentration sacran aqueous solution, sacran molecules are said to be spherical as shown schematically in Fig. \ref{mechanism1}(a). At higher concentration they will extend to soft filamentous structures as in Fig. \ref{mechanism1}(b).\cite{39} When the concentration is increased more, sacran molecules will become rod-like organization by self-organization.\cite{10} Here at the concentration just below the rod-like organization, some liquid crystal precursor may be formed (Fig. \ref{mechanism1}(c)) if there are impurities or a sacran cation molecules. Sacran as an ampholyte has both cation and anion types in aqueous solution.\cite{2} In aqueous solution the cation type sacran molecular chain may attract anion type sacran molecules by a Coulomb force. In this way the sacran molecular chains nucleate in a rod-like structure. These rod-like crystalline precursors are suggested to correspond to the 2 \textmu m-sized area seen with the highest intensity SHG signal in the center of the SHG spots in Fig. \ref{polarization}. Figure \ref{polarization}(c) also shows a clear incident polarization dependence of the SHG signal at the center (Area 1). It means that molecules are well oriented in the liquid crystal precursor. At higher sacran concentration, more liquid crystal layers are formed with various molecular orientations. However, the reason for the different molecular orientations between different layers is not clear. 

Considering the SHG images of the films in Fig. \ref{film}(c), in contrast to the SHG spots seen in the Figs. \ref{lump_and_fiber}(c), \ref{lump_and_fiber}(f), \ref{filter}(c) and \ref{polarization}(a), the SHG signal of the film is more continuous and intensity variation is small except for several dark regions of tens of micrometers in diameter. Actually, this continuous SHG was not observed in the entire range of the cast film, but only on the area placed several hundred micrometers from the edge contour of the film. Since there must be non-centrosymmetric structure in order that this SHG takes place, the sacran molecules in this part of the films must be oriented.

\begin{figure}[h]
\begin{minipage}[t]{8.5cm}
\resizebox{1\textwidth}{!}{  \includegraphics{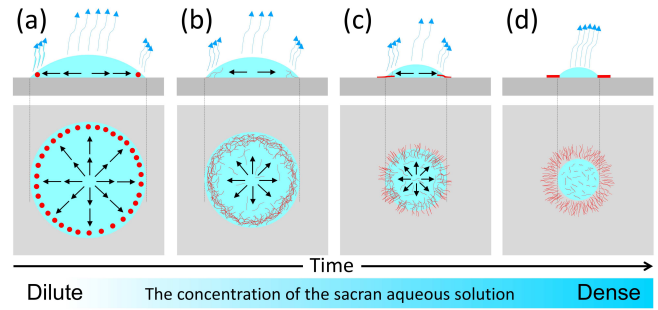}}
\end{minipage}%
 \caption{One candidate mechanism of the orientation of sacran molecules after drying from an aqueous solution droplet on a Si substrate.}
\label{mechanism2} 
\end{figure}

As one of the candidate origins of the oriented sacran molecules in the cast film, we propose the so-called "coffee-ring effect".\cite{40,41} Soon after the aqueous solution was dropped on a silicon wafer, the water evaporates fastest near the edge of the droplet, since the surface area of the droplet is large due to the curvature. Since the sacran concentration in the aqueous solution is not uniform, the gradient of the surface tension in the droplet is not uniform, either. This results in the solution flow toward the edge by Marangoni convection. Hence sacran molecules in the solution will be pulled to the edge of the droplet (Fig. \ref{mechanism2}(a)). On the other hand, when the concentration reaches a certain value, the state of sacran molecules will change from spherical into rod-like organization (Fig. \ref{mechanism2}(b)).\cite{39} In the flowing liquid when the spherical sacran molecules change into rod-like sacran, the drag on the molecules increases and they accelerate and become oriented (Fig. \ref{mechanism2}(c)). Filamentous sacran molecules may be directed towards the center of the film (Fig. \ref{mechanism2}(d)), acquire macroscopic asymmetry, and hence become SHG active. The origin of the holes of several tens of micrometers in the SHG images (Fig. \ref{film}(c)) is unknown. 

Another candidate origin of the SHG in Fig. \ref{film}(c) might be that the molecular orientation occurs simply at a certain thickness of the films. Generally, films produced by a method like in this paper take pan-cake shapes with hollows at their centers.\cite{42,43,44} The films of sacran prepared in this study are thinner (about 5 \textmu m) in the center and thicker (about 15 \textmu m) at the edges. A certain thickness may be required in order for the liquid crystal domains of sacran molecules to be formed, and that may have led to the observed SHG in Fig. \ref{film}(c).

\section{Summary}
We have observed SHG from sacran aggregates for the first time. Sacran was in the forms of ‘cotton’ lump and films. SHG spots of several tens of micrometer size were observed from pure cotton-like sacran, a sacran fiber and thin film sacran. The origin of the SHG spots could be sacran cations surrounded by anions or liquid crystal precursor due to impurities or large sacran molecules because it disappeared when the films were produced using membrane filtration of the source sacran aqueous solution. We also found a dependence of the SHG images on the polarization of the incident light. This indicates that sacran molecules in aggregates have anisotropic structure. The polarization dependent SHG microscopic images also showed multilayer structure of liquid crystal domains, and it means that each domain structure has its own orientation. On the other hand, in the film made from sacran aqueous solution, more continuous SHG was observed near the edges of the films. One of the candidate origins of this more continuous SHG is a non-uniform concentration distribution of sacran in the films caused by different evaporation velocity of water from the solution droplet on the substrate at the stage of sample preparation.

\bibliographystyle{0.zidingyi}
\bibliography{sacran_josaa_refs}
\end{document}